\documentclass[12pt]{article}
\usepackage[dvipdfmx]{graphicx}
\usepackage{here}
\newcommand{\beq}{\begin{equation}}
\newcommand{\eeq}{\end{equation}}
\newcommand{\bea}{\begin{eqnarray}}
\newcommand{\eea}{\end{eqnarray}}

\def\e2sig{e^{-2r\sigma}}

\begin{document}
\setlength{\baselineskip}{18pt}

\begin{titlepage}

\begin{flushright}
OCU-PHYS 445
\end{flushright}

\vspace*{10mm}

\begin{center}
{\bf\Large Anomalous Top and Bottom Yukawa Couplings \\
\vspace*{2mm}
and LHC Run 1 Data}
\end{center}

\vspace*{10mm}
\begin{center}
{\Large Nobuhito Maru$^{~a}$} {\large and}  {\Large Nobuchika Okada$^{~b}$}
\end{center}
\vspace*{0.2cm}
\begin{center}
${}^{a}${\it Department of Mathematics and Physics, Osaka City University, \\
Osaka 558-8585, Japan}
\\[0.2cm]
${}^{b}${\it 
Department of Physics and Astronomy, University of Alabama, \\
Tuscaloosa, Alabama 35487, USA} 
\end{center}
\vspace*{2cm}
\begin{abstract}
In some extensions of the Standard Model, 
  Yukawa couplings of the physical Higgs boson can be deviated from those in the Standard Model.  
We study a possibility whether or not such anomalous Yukawa couplings are consistent with the LHC Run 1 data. 
It is found that sizable deviations of top and bottom (and tau) Yukawa couplings 
  from the Standard Model predictions can nicely fit the data. 
New physics beyond the Standard Model can be revealed 
  through more precise measurements of such anomalous Yukawa couplings 
  at the LHC Run 2 in the near future. 
We also discuss a simple setup which can leads to anomalous Yukawa couplings.  
\end{abstract}
\end{titlepage}

After the discovery of Higgs boson at the Large Hadron Collider (LHC) \cite{ATLAS, CMS}, 
the next task of the LHC experiment is to test 
whether the Higgs boson properties are those predicted by the Standard Model (SM) or not. 
Among various data of testing the consistency of the SM predictions or searching for new physics beyond the SM, 
we focus on the Higgs boson coupling to the SM fermions, {\em i.e.} Yukawa couplings.

In the SM, a Yukawa coupling is given by
\bea
Y_f \bar{f} \Phi f = Y_f \bar{f} \langle \Phi \rangle f + \frac{Y_f}{\sqrt{2}} \bar{f} H f, 
\eea
   where $\Phi$ is the electric charge neutral component of the SM Higgs doublet field, 
   $H$ is the physical Higgs boson obtained by expanding $\Phi$ around its vacuum expectation value (VEV), 
   and $f$ represents an SM fermion.   
In the right-hand side, the first term is a fermion mass $m_f = Y_f \langle \Phi \rangle= Y_f v/\sqrt{2}$ 
  with $v\simeq 246$ GeV, and 
   the second term denotes the Yukawa coupling between the physical Higgs boson and the SM fermion. 
In the SM, the Yukawa coupling of the physical Higgs boson with the fermion is predicted 
    to be the ratio of the fermion mass to the Higgs VEV, $Y_f=m_f/v$. 
However, this fact is not necessarily true in some extensions of the SM model. 
In the two Higgs doublet model \cite{Branco} or some extensions of the Randall-Sundrum (RS) model \cite{RS}, 
  for instance, it is known that the Yukawa couplings of the physical Higgs boson 
  can be generically deviated from the SM predictions.

In this short paper, we study a possibility whether such anomalous Yukawa couplings are consistent 
   with the LHC Run 1 data \cite{LHCdata}.  
We consider deviations only in the third generation fermions 
  since FCNC constraints for the first and second generations are very severe. 
For example, if the top Yukawa coupling is suppressed, then Higgs production via the gluon fusion is decreased. 
If the bottom Yukawa coupling is also suppressed, then the decay rate of Higgs to $b\bar{b}$ is decreased, 
but other decay rates are relatively increased. 
As a result, the signal strength of the Higgs boson decay modes except for $H \to b \bar{b}$ 
  can remain almost the same as the SM predictions. 
We numerically analyze the effect of anomalous top and bottom (and tau) Yukawa couplings 
  on the signal strengths of Higgs boson decay modes for  $H \to \gamma\gamma, WW, ZZ, b\bar{b}, \tau \bar{\tau}$,  
  and examine their consistency with the LHC Run 1 data.

Parametrizing the deviations of top and bottom Yukawa couplings from the SM ones as
\beq
Y_t = c_t Y_t^{{\rm SM}}, \quad  Y_b = c_b Y_b^{{\rm SM}},
\eeq 
the partial decay widths for various modes are given by
\bea
\Gamma_{H \to gg} &=& c_t^2 \Gamma_{H \to gg}^{{\rm SM}}, \\
\Gamma_{H \to bb} &=& c_b^2 \Gamma_{H \to bb}^{{\rm SM}}, \\
\Gamma_{H \to \gamma\gamma} 
&=& \frac{\alpha_{em}^2 m_H^3}{256\pi^3 v^2} 
\left[ \frac{4}{3} c_t^2 F_{1/2}(m_H) + F_1(m_H) \right]^2, \\
\Gamma_{H \to \gamma Z} &=& 
\frac{\alpha_{em}m_W^2 m_H^3}{128\pi^4 v^4} \left( 1 - \left( \frac{m_Z}{m_H} \right)^2\right)^3 
\left[ \frac{2}{\cos \theta_W} \left( 1- \frac{8}{3} \sin^2 \theta_W \right)c_t F_{1/2}(m_H) + F_1(m_H) \right]^2, \\
\Gamma_{H \to cc} &=& \Gamma_{H \to cc}^{{\rm SM}}, \\
\Gamma_{H \to \tau\tau} &=& \Gamma_{H \to \tau\tau}^{{\rm SM}}, \\
\Gamma_{H \to WW} &=& \Gamma_{H \to WW}^{{\rm SM}}, \\
\Gamma_{H \to ZZ} &=& \Gamma_{H \to ZZ}^{{\rm SM}}, \\
\Gamma_{{\rm total}} &=& \Gamma_{H \to gg} + \Gamma_{H \to bb} + \Gamma_{H \to \gamma\gamma} 
+ \Gamma_{H \to \gamma Z} 
+ \Gamma_{H \to cc} + \Gamma_{H \to \tau\tau} 
+ \Gamma_{H \to WW} + \Gamma_{H \to ZZ}, 
\eea
where the partial decay widths in the SM are given by 
\bea
\Gamma_{H \to gg}^{{\rm SM}} &=& \frac{\alpha_s^2 m_H^3}{128\pi^3 v^2} (F_{1/2}(m_H))^2, \\
\Gamma_{H \to WW}^{{\rm SM}} &=& \frac{3m_W^4 m_H}{32\pi^3 v^4} G\left( \frac{m_W^2}{m_H^2} \right), \\
\Gamma_{H \to ZZ}^{{\rm SM}} &=& \frac{3m_Z^4 m_H}{32\pi^3 v^4} 
\left( \frac{7}{12} -\frac{10}{9} \sin^2 \theta_W + \frac{40}{9} \sin^4 \theta_W \right)
G\left( \frac{m_W^2}{m_H^2} \right), \\
\Gamma_{H \to bb}^{{\rm SM}} &=& \frac{3m_H m_b^2}{8\pi v^2} \left( 1- \frac{4m_b^2}{m_H^2} \right)^{3/2}, \\
\Gamma_{H \to cc}^{{\rm SM}} &=& \frac{3m_H m_c^2}{8\pi v^2} \left( 1- \frac{4m_c^2}{m_H^2} \right)^{3/2}, \\
\Gamma_{H \to \tau\tau}^{{\rm SM}} &=& \frac{m_H m_\tau^2}{8\pi v^2} \left( 1- \frac{4m_\tau^2}{m_H^2} \right)^{3/2}, 
\eea
and the loop functions are \cite{guide} 
\bea
F_{1/2}(m_H) &=& -2 \frac{4m_t^2}{m_H^2} \left[ 1- \left( 1 - \frac{4m_t^2}{m_H^2} 
{\rm arcsin}^2\left( \frac{m_H}{2m_t} \right) \right) \right], \\
F_1(m_H) &=& 2 + 3 \frac{4m_W^2}{m_H^2} + 3\frac{4m_W^2}{m_H^2} \left( 2 -\frac{4m_W^2}{m_H^2} 
{\rm arcsin}^2\left( \frac{m_H}{2m_t} \right) \right), \\
G(x) &=& 3 \frac{1-8x+20x^2}{\sqrt{4x-1}} {\rm arccos}\left( \frac{3x-1}{2x^{3/2}} \right) 
- \frac{1-x}{2x}(2-13x+47x^2) \nonumber \\
&& -\frac{3}{2} (1-6x+4x^2)\log x. 
\eea
In our analysis, we employ the center value of the Higgs boson mass $m_H=125.09$ GeV \cite{LHCdata}
   and the center value of the combination of Tevatron and LHC measurements 
   of top quark mass $m_t=173.34$ in GeV \cite{Mt}.

Signal strength ratios of the Higgs boson production via mainly gluon fusion and its decay are calculated from
\bea
\mu^{\gamma \gamma}
&=& \frac{\Gamma_{H\to gg} + 0.11~\Gamma^{SM}_{H \to gg}}{\Gamma^{SM}_{H \to gg} + 0.11~\Gamma^{SM}_{H \to gg}} 
\times \frac{BR(H \to \gamma\gamma)}{BR(H \to \gamma\gamma)_{SM}}, \\
\mu^{WW}
&=& \frac{\Gamma_{H\to gg} + 0.11~ \Gamma^{SM}_{H \to gg}}{\Gamma^{SM}_{H \to gg}+0.11~\Gamma^{SM}_{H \to gg}} 
\times \frac{BR(H \to WW)}{BR(H \to WW)_{SM}}, \\
\mu^{ZZ}&=& \frac{\Gamma_{H\to gg} + 0.11~ \Gamma^{SM}_{H \to gg}}{\Gamma^{SM}_{H \to gg} + 0.11~\Gamma^{SM}_{H \to gg}} 
\times \frac{BR(H \to ZZ)}{BR(H \to ZZ)_{SM}}, \\ 
\mu^{bb} &=& \frac{\Gamma_{H\to gg} + 0.11~ \Gamma^{SM}_{H \to gg}}{\Gamma^{SM}_{H \to gg} + 0.11~\Gamma^{SM}_{H \to gg}} 
\times \frac{BR(H \to bb)}{BR(H \to bb)_{SM}}, \\
\mu^{\tau\tau} &=& \frac{\Gamma_{H\to gg} + 0.11~ \Gamma^{SM}_{H \to gg}}{\Gamma^{SM}_{H \to gg} 
+ 0.11~\Gamma^{SM}_{H \to gg}} 
\times \frac{BR(H \to \tau\tau)}{BR(H \to \tau\tau)_{SM}}.  
\eea
Here, we have added $0.11 \Gamma^{{\rm SM}}_{H \to gg}$ 
   which corresponds to the contributions of Higgs boson production by the vector boson fusion channels, 
  and other contributions to Higgs boson production have been neglected (see Table~7 in \cite{LHCdata}). 
%
  
\begin{figure}[t]
\begin{center}
\includegraphics[width=10.0cm,bb=0 0 298 294]{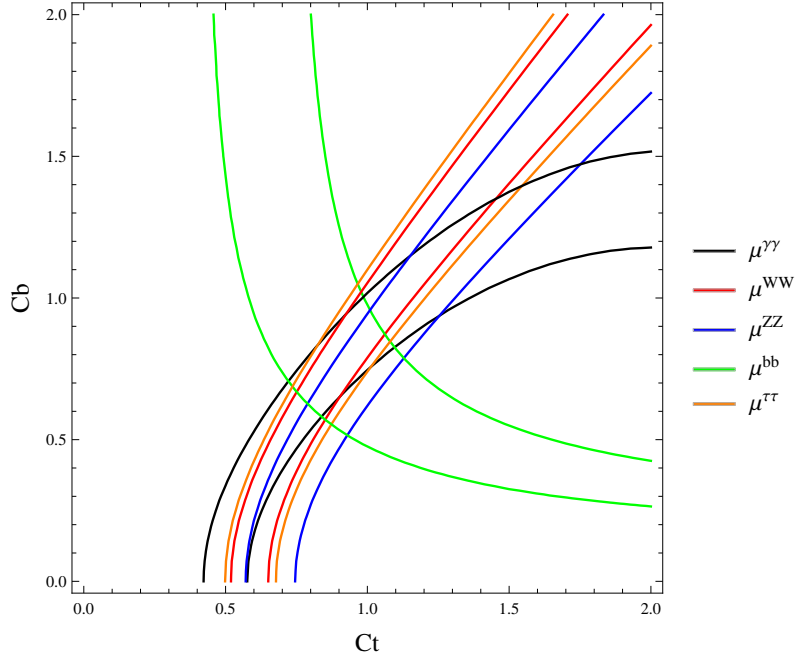}
\end{center}
\vspace*{-7mm}
\caption{
Signal strength ratios in $(c_t, c_b)$-plane. 
Here, we have used the constraints from the LHC Run 1 data for various modes (68\% confidence level): 
 $0.98  \leq  \mu^{\gamma \gamma} \leq 1.36$, 
 $0.94  \leq  \mu^{WW}  \leq1.29$, 
 $1.07  \leq  \mu^{ZZ}    \leq1.58$, 
 $0.42  \leq  \mu^{bb}     \leq 0.98$, and 
 $0.89 \leq \mu^{\tau \tau} \leq1.37$. 
}
\label{result1}
\end{figure}

Fig.~\ref{result1} shows our results in $(c_t, c_b)$-plane for various signal strength ratios 
   set by the LHC Run 1 data at 68\% confidence level \cite{LHCdata}. 
Interestingly, we can see that suppressed top and bottom Yukawa couplings, 
  $c_t \simeq 0.9$ and $c_b\simeq 0.8$, nicely fit the LHC Run 1 data \cite{LHCdata}.

\begin{figure}[t]
\begin{center}
\includegraphics[width=10.0cm,bb=0 0 298 294]{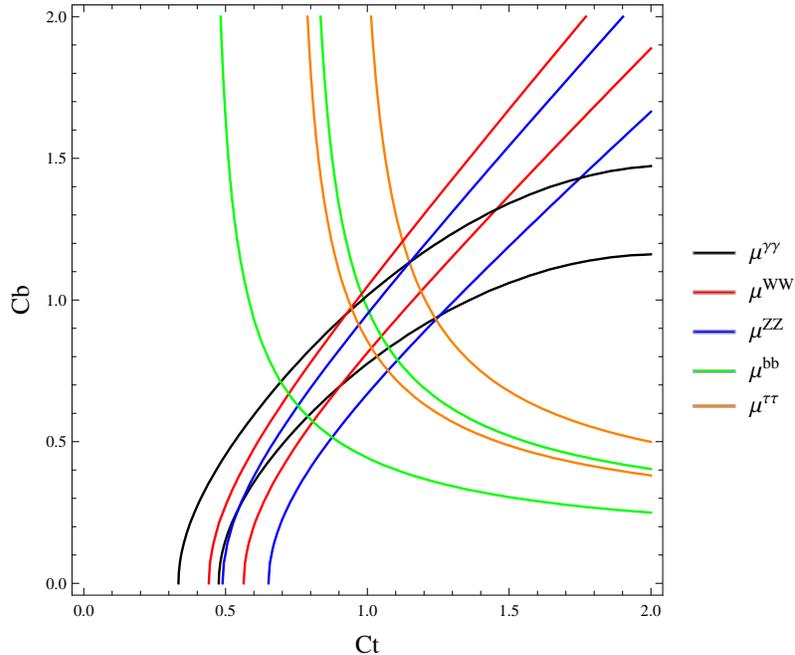}
\end{center}
\vspace*{-7mm}
\caption{
Same as Fig.~\ref{result1}, but we have here imposed the GUT relation of $c_b=c_\tau$. 
}
\label{GUT}
\end{figure}

It may be interesting to consider a unification of bottom and tau Yukawa couplings, 
  which is usually predicted in Grand Unified Theory (GUT). 
Although this relation holds only at a GUT scale, we apply it to our analysis, for simplicity. 
See, for example, \cite{TeVGUT} for a GUT scenario with the coupling unification at the TeV scale.   
Let us introduce an anomalous coupling to tau Yukawa coupling, and parameterize it as 
\bea 
  Y_\tau=c_\tau Y^{\rm SM}_\tau,  
\eea
   and hence $\Gamma_{H \to \tau \tau} = c_\tau^2 \Gamma_{H \to \tau \tau}^{{\rm SM}}$.   
By imposing the GUT relation of $c_b=c_\tau$, we calculate signal strength ratios. 
Our results are shown in Fig.~\ref{GUT}. 
We find that $c_t \simeq 1$ and a slightly suppressed bottom Yukawa coupling $c_b\simeq 0.9$ 
  fit the LHC Run 1 data \cite{LHCdata}.

Finally, we briefly discuss the five dimensional SM in the Randall-Sundrum background (see, for instance \cite{CHNOY}), 
   where the physical Higgs boson couplings to the SM fermions 
   are indeed deviated from the SM prediction. 
The original RS model \cite{RS} was proposed to provide a solution to the gauge hierarchy problem 
   by using five dimensional AdS spacetime with a metric 
\bea
ds^2 &=& G_{MN} dx^M dx^N = e^{-2ky} \eta_{\mu\nu} dx^\mu dx^\nu - dy^2, \\
\eta_{\mu\nu} &=& {\rm diag}(+,-,-,-,-), 
\eea
compactified on an orbifold $S^1/Z_2$. 
Here, $x, y$ are the coordinates of four dimensional spacetime and the fifth component, 
and $M,N (\mu, \nu)$ denote the indices for the five (four) dimensional spacetime.  

Suppose that the SM Higgs field $H(x,y)$ propagates in the bulk with the action  
\bea
S_H= \int d^4x dy \sqrt{-G} \left[ G^{MN} \partial_M H^\dag \partial_N H - \lambda (H^\dag H)^2 + m^2 H^\dag H \right], 
\eea
where the ordinary quartic Higgs potential is assumed. 
The equation of motion is obtained as
\bea
0 = -e^{2ky} \partial_\mu \partial^\mu H - \partial_y \partial^y H + 4k \partial^y H -(2\lambda H^\dag H - m^2) H.  
\eea
It is easy to see that the constant VEV is a solution of the equation of motion, $v=\sqrt{m^2/(2\lambda)}$. 
Expanding the Higgs field around the VEV in terms of Kaluza-Klein mode functions $f_n(y)$ as 
$H(x,y) = v + \tilde{H}(x,y)$ and $\tilde{H}(x,y) = \sum_n f_n(y) \tilde{H}_n(x)$
and plugging it back into the equation of motion, we find
\bea
\frac{d^2 f_n(y)}{dy^2} -4k \frac{d f_n(y)}{dy} - ( e^{2ky} m_n^2 - 2m^2) f_n = 0
\eea
where $- \partial_\mu \partial^\mu \tilde{H}(x) = m_n^2 \tilde{H}(x)$, and we have neglected the nonlinear term. 

The solution of this equation is given by Bessel functions, $J_\nu$ and $Y_\nu$, as 
\bea
f_n(y) = \frac{e^{2ky}}{\sqrt{N_n}} \left[ J_\nu(m_n e^{ky}/k) + b_\nu(m_n) Y_\nu(m_n e^{ky}/k) \right]
\eea
where $\nu=\sqrt{4+2m^2/k^2}$, $N_n$ is a normalization constant, and 
  $b_\nu(m_n)$ is a $y$ independent function determined by the boundary conditions. 
Now, our interest is on the lightest Kaluza-Klein mode ($n=1$). 
Noting that the Bessel functions are approximated for $m_1 e^{ky}/k \ll 1$ as follows, 
\bea
&&J_\nu (m_1 e^{ky}/k)  \simeq \left( \frac{m_1 e^{ky}}{2k} \right)^\nu \frac{1}{\Gamma(\nu+1)}, \\
&&Y_\nu (m_1 e^{ky}/k)  \simeq -J_{-\nu}(m_1 e^{ky}/k) \simeq  \left( \frac{m_1 e^{ky}}{2k} \right)^{-\nu} \frac{1}{\Gamma(-\nu+1)},
\eea
then, the mode function for the physical Higgs boson $f_1(y)$ has a nontrivial $y$-dependence,
while the Higgs VEV is a constant.  

Quarks and leptons are also bulk fields and can have nontrivial profiles in the fifth dimension. 
The four dimensional effective Yukawa couplings are given by the overlap integral 
of the quark and lepton mode functions and the mode function for the physical Higgs field. 
Therefore, Yukawa couplings between the physical Higgs boson and fermions are deviated from the SM ones.

In summary, 
we have studied anomalous Yukawa couplings motivated by the fact that 
  the relation between the Yukawa couplings of the physical Higgs boson 
  and the fermion mass in the SM can be violated 
  in some extensions of the SM, such as two Higgs doublet models and Randall-Sundrum type models. 
Focusing on the top and bottom Yukawa couplings, 
  we have shown that the anomalous top and bottom Yukawa couplings can fit 
  the LHC Run 1 data for the signal strength at a variety of Higgs boson decay modes. 
We have also considered the GUT extension case where the bottom and tau Yukawa couplings 
  are unified at the GUT scale.  
Even in this case, the deviations of the Yukawa couplings from the SM predictions are 
  found to be consistent with the LHC Run 1 data. 

The anomalous Yukawa couplings will be more precisely tested by the LHC Run 2 and High-Luminosity LHC in the near future. 
New physics beyond the SM may be revealed through deviations of the Higgs boson couplings to SM fermions.

\subsection*{Acknowledgments}
We would like to thank Y. Seiya and K. Yamamoto for fruitful discussions.  
The work of N.O. is supported in part by the United States Department of Energy grant (DE-SC0013680).


\end{document}